\documentclass[12pt]{article}

\usepackage{epsfig}
\usepackage{inputenc}
\def\htmladdnormallink#1#2{#1}
\inputencoding{latin1}

\def\sub#1{_{\mbox{\scriptsize #1}}}
\def\m#1{\mbox{#1}}
\def\code#1{\textbf{\texttt{#1}}}
\def\event{\code{Event}}
\def\collision{\code{Coll\-ision}}
\def\step{\code{Step}}

\def\decayer{\code{Decayer}}
\def\selector{\code{Selector}}
\def\decayrater{\code{Decay\-Rater}}
\def\subcollision{\code{Sub\-Collision}}
\def\particledata{\code{Particle\-Data}}
\def\particlematcher{\code{Particle\-Matcher}}
\def\stop{\code{Stop}}
\def\veto{\code{Veto}}
\def\mihandler{\code{Multiple\-Interaction\-Handler}}
\def\subhandler{\code{Sub\-Pro\-cess\-Handler}}
\def\cascadehandler{\code{Cascade\-Handler}}
\def\hadronizationhandler{\code{Hadronization\-Handler}}
\def\decayhandler{\code{Decay\-Han\-dler}}
\def\lumifn{\code{Luminosity\-Function}}
\def\partonextractor{\code{Parton\-Extractor}}
\def\partondensity{\code{Parton\-Density}}
\def\partonxsec{\code{Parton\-XSec\-Fn}}
\def\hint{\code{Hint}}
\def\stephandler{\code{Step\-Handler}}
\def\eventhandler{\code{Event\-Handler}}

\def\collisionhandler{\code{Collision\-Handler}}
\def\sigmaxfn{\code{SigmaMaxFn}}
\def\noll{\code{null}}
\def\remnanthandler{\code{Remnant\-Handler}}
\def\particle{\code{Particle}}
\def\interface{\code{Interface}}
\def\shat{$\hat{s}$}
\def\cpp{{\sc C++}}
\def\ariadne{{\sc Ariadne}}

\def\jetset{{\sc Jetset}}
\def\pythia{{\sc Pythia}}
\def\pyth{{\sc Pythia7}}

\def\hepevt{\code{HEPEVT}}
\def\lujets{\code{LUJETS}}

\def\laeq{\,\lower3pt\hbox{$\buildrel < \over\sim$}\,}

\def\qq{\mbox{q}\overline{\mbox{q}}}

\newcounter{Aenumct}

\begin{document}

\begin{titlepage}

  \renewcommand{\thefootnote}{\fnsymbol{footnote}}

  \begin{flushright}
    LU--TP 98--21\\
    hep-ph/9810208\\
    October 1998
  \end{flushright}
  \begin{center}

    \vskip 10mm
    {\LARGE\bf Development Strategies for \pythia\ version 7}
    \vskip 15mm

    {\large Leif Lönnblad}\\
    Department of Theoretical Physics\\
    Sölvegatan 14a\\
    S-223 62  Lund, Sweden\\
    leif@thep.lu.se

  \end{center}
  \vskip 50mm
  \begin{abstract}
    
    \hskip -3mm This document describes the strategies for the
    development of the \pyth\ program. Both the internal and external
    structure of the program is discussed. Some comments on
    relationship to other software is given as well as some comments
    on coding conventions and other technical details.

\end{abstract}

\end{titlepage}

\tableofcontents

\newpage
 
\section{Introduction}

\pyth\ \cite{pyt7} will be a new event generator well suited to meet
the needs of future high-energy physics projects, for phenomenological
and experimental studies. The main target is the LHC community, but it
will work equally well for linear e$^+$e$^-$ colliders, muon
colliders, the upgraded Tevatron, and so on.  The generator will be
based on the existing Lund program family, but rewritten from scratch
in a modern, object-oriented style, using \cpp. The greatly enhanced
structure will make for improved ease of use, extendibility to new
physics aspects, and compatibility with other software likely to be
used in the future.

The current state-of-the-art event generator programs --- \pythia,
\jetset\ and \ariadne\ for the purposes of this application ---
generally work well. It has been possible gradually to extend them
well beyond what could originally have been foreseen, and thus to
parallel the development of the high-energy physics field as a whole
towards ever more complex analyses. However, a limit is now being
approached, where a radical revision is necessary, both of the
underlying structure and of the user interface.

Even more importantly, there is a change of programming paradigm,
towards object-oriented methodology. In the past, particle physicists
have used Fortran, but now \cpp\ is taking over. \cpp\ has been
adopted by CERN as the main language for the LHC era. The CERN program
library is partly going to be replaced by commercial products, partly
be rewritten in \cpp.  In particular, the rewriting of the detector
simulation program Geant \cite{Geant4} is a major ongoing project,
involving several full-time programmers for many years, plus voluntary
efforts from a larger community.  The CERN shift is matched by
corresponding decisions elsewhere in the world: at SLAC for the
B-factory, at Fermilab for the Tevatron Run 2, and so on. \cpp\ is
also the language adopted by industry (plus Java, for Internet
applications).
   
Therefore a completely new version of the Lund programs, written in
\cpp, is urgently called for. In order to reap the benefits of the
\cpp\ language, it is not enough with a mechanical conversion (like
what f2c offers from Fortran to C). What is required is a complete
rethinking of the way an event generator should look.

This paper will lay out the development strategies for the \pyth\ 
event generator. It will focus on the basic structure for the event
representation and the event generation, but will also discuss other
aspects of the development, such as the relation between parts of the
\pyth\ code and other High Energy Physics (HEP) software, the user
interface and documentation. In addition, in the appendix some coding
conventions will be discussed together with the licensing policy.

This paper is not the final word on the structure of \pyth, but is
meant to initiate a discussion between people interested in the future
development of \pyth\ and event generation in general. Comment and
suggestions are more than welcome.

\section{The structure of Event Generation}

To discuss the structure of event generation it is, of course,
reasonable to first look at the structure of an \emph{event}.
Experimentally, an event is some combinations of hits in a detector,
triggering the recording of all hits present in a given time interval.
A computer generated event is different. Since detector simulation is
not normally included, it does not contain hits, but rather a set of
particles which could give hits in the detector, thus possibly
triggering a recording. But it may also contain other information,
which could be important for the analysis of the generated event.

It is instructive to see what is provided by conventional event
generators via eg.\ the \lujets\ \cite{jetset} or \hepevt\ 
\cite{lepws} common blocks. These are essentially numbered lists of
particle entries, but they also contains a lot of information about
how the event was generated. For example, the entry of a particle
which has decayed includes numbers representing the range of entries
containing its decay products, and vice versa, each of the decay
product entries has a number representing the entry of its parent.
The common blocks may also include entries describing one or several
hard sub-processes, and may also contain particles from several
different collisions which could all be registered as one event.

It is clear that object orientation enables us to have the actual
structure of the \emph{event object} to reflect the history of the
event generation in a much more natural way. This structure may become
very complex, but it is important to keep in mind that, while doing
complex analysis of a generated event requires a complex structure, it
must still be very simple to do a simple analysis. This is
accomplished in \cpp\ using the \textit{data hiding} aspect of object
orientation, where the user need not know about the way data is
actually structured inside an object, but only about the methods for
extracting the desired information.

\begin{figure}[t]
  \begin{center}
    \epsfig{figure=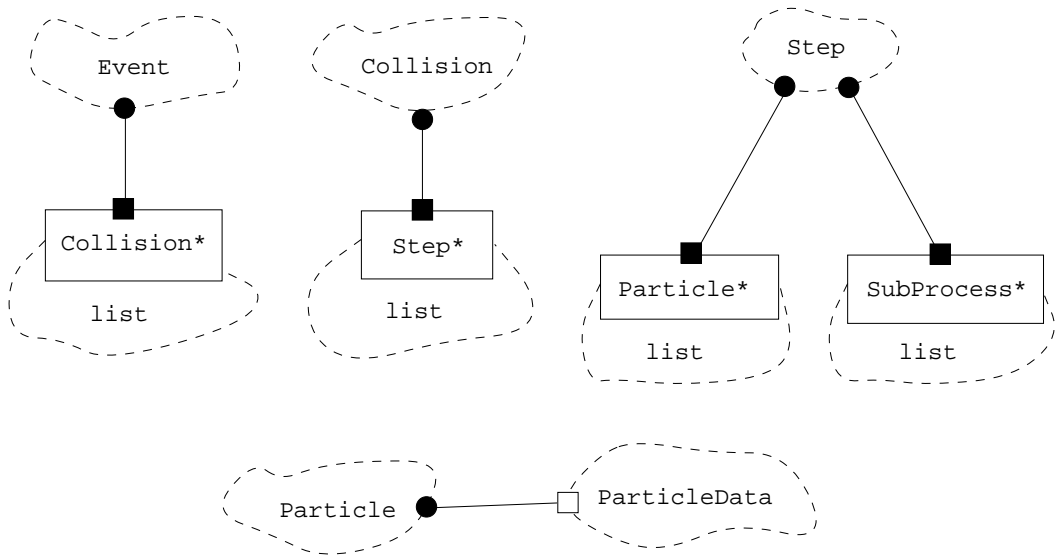,width=14cm}
    \caption[dummy]{{\it Class diagram for the suggested structure of
        a generated event in \pyth. (See appendix \ref{apx:booch} for
        a quick guide on Booch notation.) }}
    \label{fig:event}
  \end{center}
\end{figure}

\subsection{The structure of a generated event}
\label{sec:eventstructure}

For \pyth\ the suggested internal structure of a generated event is
given in fig.~\ref{fig:event} as a simplified class diagram in Booch
notation\cite{booch}.

To exemplify the structure we use the case of Z$^0$ production at the
LHC. The luminosity at the LHC is so high that each \event\ will
consist of several \collision s between the protons in each bunch
crossing. One \collision\ is special since it will contain the physics
process (Z$^0$ production) under study and will be called the
\emph{trigger} collision. The rest will be of minimum-bias
type and are called \emph{overlayed} collisions.

Each \collision\ consists of a number of \step s. Each step contains a
list\footnote{Note that throughout this document, \textit{list}
  represents a \emph{container} in general. In each case the
  \texttt{stl} container most suited for the task at hand will be
  used.} of outgoing particles from the collision describing the state
of the event after a given step in the event generating process. As we
shall see below, the actual steps needed to generate events are
process- and model-dependent. In this example, the first step in the
trigger collision would be the generation of the hard sub-collision
eg.\ $\qq\rightarrow\mbox{Z}^0$. The second could be the parton
cascade. After that there may be secondary sub-collisions between the
partons in the colliding protons generated, followed by the
hadronization and the decay of unstable particles. If the Z$^0$ then
decays into quarks there may be an additional parton cascade step
followed by hadronization and decay.

Here follows a more detailed description of the classes used to
describe an event.

\subsubsection*{\particledata}

A \textit{Particle} is, of course, one of the most central concepts in
HEP. It is, however, not easy to translate the general notion of a
particle to a (C++) object in a way acceptable to everyone (see
section \ref{sec:CLHEPevent} for an elaboration on this problem).

In \pyth, a \particle\ object representing an instance of a
given particle type will be clearly separated from the \particledata\ 
object representing the properties of the particle type. A
\particle\ will contain a pointer to the \particledata\ object
containing information common to all instances of the particle type,
together with information specific to the actual particle instance.

A \particledata\ object will contain the following information:
\begin{itemize}
  \itemsep -1mm
\item The particle id number as given by PDG \cite{pdg}.
\item A character string representing the particle
  name\footnote{Possibly in different formats, eg.\ \LaTeX, HTML
    etc. This will hopefully also be standardized by PDG.}
\item The nominal mass.
\item A pointer to a \code{ParticleMass} object.
\item The width and/or the average lifetime.
\item The charge.
\item The spin.
\item The colour.
\item A flag to tell if the particle is stable or not.
\item A pointer to the corresponding anti-particle.
\item A flag to tell if the corresponding anti-particle should be
  modified in parallel with the particle.
\item A decay table.
\end{itemize}

The \particledata\ class will also define an enumeration, relating
each of the standard particle codes to sensibly named identifiers,
eg.\ \code{ParticleData::piplus=211}.

The pointer to the \code{ParticleMass} object is typically \noll\ for
most particles. But for some particles it is not possible to define a
unique nominal mass. For eg.\ quarks one could define a
\emph{constituent} mass, a \emph{current algebra} mass, and a running
mass according to some renormalization scheme. Such information is
stored in the \code{ParticleMass} object.

The properties of an anti-particle can, of course, be deduced from
those of the particle. To simplify the access, however, the properties
of the particle and its anti-particle are stored as two separate
objects. Instead the properties of the anti-particle will by default
be automatically updated whenever the particle is changed, and vice
versa.  Optionally, a flag can be set to completely decouple the two.

The decay table is a list of decay modes specifications, each
associated with a decay fraction ($\Gamma_i/\Gamma$) and a \decayer\ 
object, which is able to perform the actual decay. The decay mode
specification may be a simple list of the decay products, but may also
be much more complicated. It is up to the \decayer\ object to say
whether it is able to perform the specified decay or not.

Besides the list of decay products, the specification may contain a
list of \particlematcher\ objects, where each object corresponds to
one decay product from a set of particles represented by this
matcher\footnote{This is the same scheme which is used in the HepPDT
  classes\cite{HepPDT}}. It may also contain a \particlematcher\ 
object corresponding to any number of decay products of the matching
set. In this way one can represent decay modes such as
\[
 \mbox{B}^+\rightarrow \mbox{\textit{any} D
    \textit{meson}} + \pi^+ + \pi^- + \mbox{\textit{any number of other
    mesons}}
\]
The decay mode may also contain a list of decay modes for specifying
subsequent resonant decays and a list of \particledata\ objects
corresponding to excluded intermediate resonances. Hence the following
three decay modes with the same final state decay products, can be made
distinct:
\begin{eqnarray}
  \mbox{D}^+&\rightarrow& \mbox{K}^-\pi^+\mbox{e}^+\nu_e
  \mbox{~(inclusive),}\nonumber \\
  \mbox{D}^+&\rightarrow& \overline{\mbox{K}}^{0\star} \mbox{e}^+\nu_e
  \mbox{~(with subsequent~} \overline{\mbox{K}}^{0\star}\rightarrow
  \mbox{K}^-\pi^+\mbox{),}\nonumber \\
  \mbox{D}^+&\rightarrow& \mbox{K}^-\pi^+\mbox{e}^+\nu_e
  \mbox{~(non-resonant).}\nonumber
\end{eqnarray}

For some particles, the decay fractions are given by models which
depends on many parameters. The decay table may therefore include a
pointer to a \decayrater\ object which may calculate the decay
fractions for each specified channel at initialization time, or even
for each instance of the particle type. The latter is eg.\ needed for
wide resonances, where the decay fractions are dependent on the actual
mass of a particle instance. Also for processes such as
$\m{e}^+\m{e}^-\rightarrow\m{Z}^0$ at a future linear collider, where
Z$^0$ is far off-shell above the $\m{t}\overline{\m{t}}$ threshold, it
is clear that one cannot use the same decay fractions as for an
on-shell Z$^0$ at LEP~1.

Each decay channel has a switch to temporarily disallow a decay, and
it is possible to switch on or off groups of decay channels for a
given particle with a search criteria, eg.\ \textit{switch off all
  leptonic decay modes for $\mbox{W}^+$}.  Each decay channel can be
associated with a \textit{K-factor} to temporarily boost the rate of a
channel which is under study.

For some analysis situations, it is desirable to have different decay
modes for the same particle type if it appears more than once in an
event.  For example, one may want to study
$\m{H}^0\rightarrow\m{Z}^0\m{Z}^0$, where one Z$^0$ decays
leptonically and the other hadronically.  This may then be specified
in the decay table of the higgs with the subsequent resonance decay
mechanism described above. The same can be done for
$\m{H}^0\rightarrow\m{W}^+\m{W}^-\rightarrow\m{e}\nu+\m{jets}$, but
here one could alternatively decouple the W$^+$ from the W$^-$, and
then switch off all leptonic channels for one and all hadronic for the
other.

The specification of subsequent resonance decays can also be used for
correlated decays. Specifying $\m{H}^0 \rightarrow \m{Z}^0\m{Z}^0
\rightarrow \qq \m{q}'\overline{\m{q}}'$, the decayer performing the
Higgs decay can be made responsible for generating the desired angular
correlations in the Z$^0$ decays. Alternatively, the decayer
responsible for the Z$^0$ decay may decide to model the correlations
by peeking at the decay of the sibling Z$^0$.

The set of particles used by \pyth\ is not fixed. The user can freely
introduce new ones, and also write new classes inheriting from the
\particledata\ and \particle\ classes.

\subsubsection*{\particle}

The \particle\ object contains the following information:
\begin{itemize}
  \itemsep -2mm
\item A pointer to the corresponding \particledata\ object.
\item A list of pointers to \particle s, corresponding to decay
  products. This list is empty if the particle has not decayed.
\item A list of pointers to \particle s, corresponding to parent
  particles. This list is empty if the particle has no ancestors.
\item A pointer to a \particle\ object corresponding to the same
  actual particle, but for which eg.\ the momenta has been modified in
  the the generating process, and therefore is no longer present among
  the final state \particle\ objects. There is also a pointer for the
  inverse relationship. These pointers are \noll\ if no corresponding
  object exists.
\item A pointer to an object carrying information about the colour
  connections (incoming/outgoing/neighboring colour/anti-colour) for
  this \particle. This pointer is \noll\ if the particle is colour
  singlet.
\item A pointer to an object carrying information about the spin state
  of this particle. This pointer is \noll\ if there is no such
  information available.
\item A Lorentz vector corresponding to the momentum of this \particle.
\item The mass of this instance of the particle. (The nominal mass is
  stored in the corresponding \particledata\ object.)
\item A Lorentz vector corresponding to the space--time point where
  this particle was created.
\item A Lorentz vector corresponding to the space--time distance this
  particle traveled before it decayed.
\item The proper lifetime of this particle. (The nominal lifetime
  and/or width is stored in the corresponding \particledata\ object.)
\item The scale at which this particle is resolved (to be used by
  parton cascade models).
\item A pointer to the \step\ object (see below) where this \particle\ 
  first occurred.
\end{itemize}

\subsubsection*{\step}

The \particle\ objects are collected in \emph{steps}, where each
\step\ corresponds to the state of a collision after a certain step in
the event generating process. There are two lists, one for the
particles which are in the final state and one for intermediate
particles which were introduced in the last generation step.  The
\step\ object may also contain pointers to \subcollision\ objects,
which are described below.

\subsubsection*{\subcollision}

A \subcollision\ object contains information on a (normally hard)
$2\rightarrow N$ sub-processes in the collision. It contains pointers
to the two incoming partons\footnote{\textit{Parton} in this context
  denotes not only quarks and gluons in a hadron, but may also be eg.\ 
  a photon within an electron. In all cases it is represented by a
  \particle\ object}, a list of pointers to the outgoing partons, and
a pointer to an object describing the underlying diagram used for the
generation.

\subsubsection*{\collision}

Besides an ordered list of \step s, a \collision\ object contains the
two colliding particles, and the space--time position of the collision
vertex.

\subsubsection*{\event}

An \event\ object contains pointers to \particledata\ objects
representing the particles in the colliding beams and a list of
\collision\ objects.  One of these \collision s is the \emph{trigger}
collision containing the process to be studied.

\subsubsection*{\selector}

As mentioned above, the typical user of the event record need not be
concerned with the internal structure of the \event\ class. Of course,
access the underlying structure is possible if this is needed for a
complicated analysis, but typically the user would call member
functions of the \event\ and \collision\ classes to access the desired
information. A typical task would be to extract a list of particles
from the \event\ object. This is done with the \texttt{extract}
method, where a \selector\ is specified representing the criteria on
the particles to be extracted. A typical \selector\ class could be one
which selects charged, final-state particles, or one which extracts
charmed mesons which have decayed semi-leptonically.

\subsection{The structure of the event generating process}

The actual event generation is performed by many different
\emph{handlers}, each responsible for a specific step in the
generation process. The two main ones responsible for the more
administrative tasks are the \eventhandler\ and \collisionhandler,
where the former is responsible for the administration of overlayed
collisions, while the latter deal with the generation of a single
collision.

Following the structure of the \collision\ class, the
\collisionhandler\ contains a list of \stephandler\ objects
responsible for the different steps in generation process. To be as
flexible as possible, the order in which different steps should be
performed in the generation is not predefined in \pyth. Not only may
the order be different in different physical models, it may also
differ for different processes within the same model framework.
However, to simplify things, the list of \stephandler s has some
structure so that handlers may be grouped according to the typical
time scales at which the corresponding physical processes are assumed
to take place. Typically, first a hard sub-process is chosen,
thereafter a parton cascade is allowed to develop, followed by the
hadronization process and the decay of unstable particles.

However this order need not be followed strictly, and in particular
each \stephandler\ is allowed to change the order in which subsequent
handlers are applied, and also jump back in the list. In the example
of Z$^0$ production above, the \stephandler\ responsible for the decay
of the Z$^0$ must, in the case the decay is into $\qq$, be able to
tell the \collisionhandler\ that a parton cascade should be applied,
followed by hadronization and particle decay.

To help the \stephandler s, they may be given a \hint\ object. In the
previous example, the handler responsible for the Z$^0$ decay would
not only tell the \collisionhandler\ that a parton cascade needs to be
performed, but could also give the hadronization handler a \hint\ 
containing eg.\ a list of pointers to the partons which need to be
cascaded.

\begin{figure}[t]
  \begin{center}
    \epsfig{figure=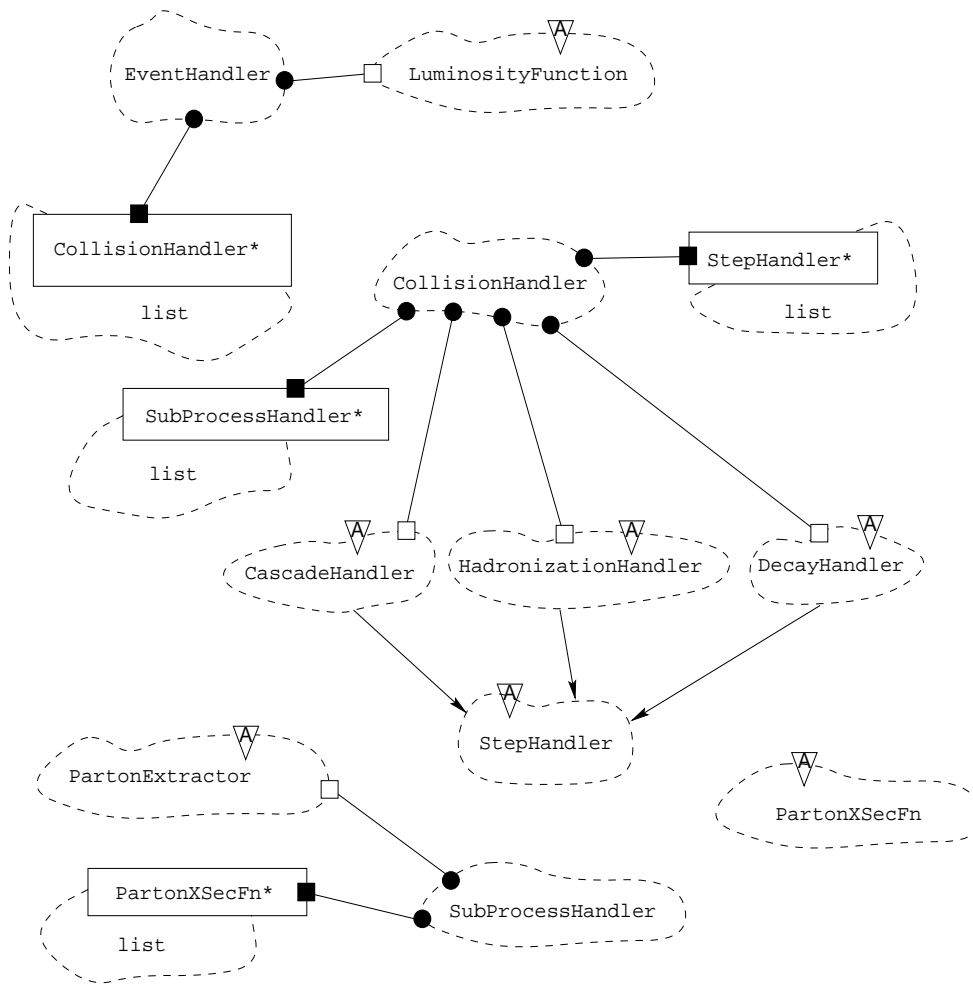,width=13cm,height=13cm}
    \caption[dummy]{{\it Class diagram for the suggested structure of
        different handlers responsible for the generating process in
        \pyth.}}
    \label{fig:handlers}
  \end{center}
\end{figure}

The structure of the handler classes is outlined in
fig.~\ref{fig:handlers} in Booch notation. Here follows a brief
description of some of the classes.

\subsubsection*{\eventhandler}

The \eventhandler\ contains a \lumifn\ object which describes the
parameters of the beams and some parameters of the detector, to be
able to estimate the number of overlayed collisions per trigger. Then
there is a \collisionhandler\ which can generate a trigger collision
and a list of other collision handlers to generate overlayed
collisions. The latter are typically of minimum-bias type and may be
collisions the beam particles as for the trigger collision, but may
also be eg.\ beam-gas collisions. The \eventhandler\ class itself is
not supposed to be inherited from, instead all extra functionality
should be implemented by inheriting from the \lumifn\ class or
configuring the \collisionhandler\ objects.

\subsubsection*{\collisionhandler}

This is by far the most complicated class among the handlers. Given a
pair of colliding particles, it will select a sub-process among a list
of \subhandler s to generate the primary sub-process. After that it
will call a number of \stephandler s divided into groups as follows
\begin{itemize}
\itemsep 0mm
\item parton cascades,
\item multiple interactions,
\item hadronization and
\item decays of unstable particles.
\end{itemize}
Each of these groups have one specialized step handler of the class
\cascadehandler, \mihandler, \hadronizationhandler\ and \decayhandler\ 
respectively, and two lists of general \stephandler s, one to be
applied before and one after each main handler. There is also a list
of \stephandler s to be applied directly after the \subhandler.

Each of the \stephandler s in any of the lists may be given a \hint\ 
object, while the cascade, multiple interaction, hadronization and
decay handlers may be given a list of \hint s. The \collisionhandler\ 
has defaults for all these handlers, but the chosen \subhandler\ may
change these for the current event. The groups of handlers above will
be traversed in order, and any handler called may add a \stephandler\ 
with a \hint\ to any of the lists, or just add a \hint\ to any of the
lists of hints for the cascade, multiple interaction, hadronization
and decay handlers.  If a handler adds a handler and a hint in a list
which the \collisionhandler\ has already gone through, the
\collisionhandler\ will afterwards restart from that group. For
example: if a decay handler gives a hint to the cascade handler to
cascade partons from a Z$^0$ decay, the \collisionhandler\ will
restart with the parton cascade group. However, each hint is only
processed once by the collision handler, even if a list is processed
several times.

Each of the groups can be individually switched on or off. If eg.\ the
cascade and multiple interaction groups have been switched off, and
the decay handler in the previous example has given a hint to the
cascade handler, the generation will then restart with the
hadronization group instead.

The \collisionhandler\ itself is not supposed to be extended with
inheritance. Instead, different models for event generation should be
implemented by inheriting from the \stephandler\ class and specifying
how they should be called and by configuring the different
\subhandler\ objects.

\subsubsection*{\subhandler}

A \subhandler\ has a \partonextractor\ object and a list of
\partonxsec\ objects. These interact so that the partonic cross
section function is convoluted with a parton-parton luminosity given
by the \partonextractor\ to get an estimate of the maximum cross
section for the given sub-process, which is necessary for the
subsequent Monte Carlo Generation. After a sub-process is selected,
the \partonxsec\ object is told to generate the kinematics of the
parton-parton scattering, with two incoming and a number of outgoing
partons. The \partonextractor\ object is then responsible for the
generation of remnants of the incoming particles.  The \subhandler\ 
also contains an object describing the kinematical cuts to be applied
to the parton-parton sub-process.

The \subhandler\ itself is not supposed to be extended with
inheritance. Instead, new sub-processes and parton density functions
are implemented by creating new classes inheriting from
\partonextractor\ and \partonxsec.

\subsubsection*{\partonextractor}

\partonextractor\ is an abstract base class, defining the methods
needed to convolute a given partonic cross-section with the
parton-parton luminosity and to generate the remnants when extracting
partons from the colliding particles. In the example of Z$^0$
production at the LHC above, the class derived from \partonextractor\ 
would typically contain a pointer to a \partondensity\ object
encapsulating the proton structure function parameterization, and a
pointer to a \remnanthandler\ (which both may be changed by the user).
As a more complicated case, we may consider collisions between
quasi-real photons in e$^+$e$^-$ colliders, where the photons may
fluctuate into vector mesons. Here the parton extractor could contain
a \partondensity\ object describing the parton content of a photon,
which would then be used in the double convolution of the partonic
cross section with the parton densities of the photons and the
spectrum of photons radiated from the electrons. The \remnanthandler\ 
would then be responsible for generating the electron and photon
remnants given the extracted photon and parton.

\subsubsection*{\partonxsec}

Also \partonxsec\ is an abstract base class, declaring methods for
returning a \sigmaxfn\ object encapsulating the estimated upper bound
on the partonic cross section as a function of the squared
parton-parton invariant mass, \shat, and for constructing the
kinematics of the parton-parton interaction. To enable the
\partonextractor\ to efficiently convolute the partonic cross section
with the parton densities, the \sigmaxfn\ is not only a simple
function object, but has a predefined form:
\begin{eqnarray}
  \label{eq:sighatfn}
  \hat{\sigma}_{\max}(\hat{s}) = c_0 + \frac{c_1}{\hat{s}} +
  \frac{c_2}{\hat{s}^2} &+& \frac{c_3}{\hat{s}(\hat{s}+r)} +
  \frac{c_4}{(\hat{s}-m^2)^2+m^2\Gamma^2}\nonumber\\ &+&
         \frac{c_5}{\hat{s}(\hat{s}+r')} +
  \frac{c_6}{(\hat{s}-m'^2)^2+m'^2\Gamma'^2}.
\end{eqnarray}
This should be sufficient to describe most partonic cross
sections. But neither the \partonextractor\ or the \partonxsec\ are
required to use this internal structure. The \partonextractor\ may use
it as a normal function object, and the \partonxsec\ may use a derived
class, setting all constants in eq.~(\ref{eq:sighatfn}) to zero and
override the method returning the function value. It is also possible
for the \partonextractor\ and \partonxsec\ classes to subdivide the
\shat\ interval, to allow for more efficient generation.

\subsubsection*{\stephandler}

The nature of a \stephandler\ may be quite varied. Besides the four
standard step handlers, the \cascadehandler, \mihandler,
\hadronizationhandler\ and \decayhandler, there may be handlers which
performs eg.\ colour reconnections or Bose--Einstein reweighting of
the event. Any \stephandler\ may throw a \veto\ exception, telling the
\collisionhandler\ to discard the current collision and generate a new
one. This means that various cuts may be implemented as \stephandler
s. A \stephandler\ may also throw a \stop\ exception so that the
generation can be stopped and later restarted by the controlling
program. Also event analysis may be implemented as a step handler.

When asked to handle a step, the \stephandler\ is passed a pointer to
the current \step\ in the generation together with a \hint\ object and
a pointer to the \collisionhandler\ from which the request originated.

\subsubsection*{\hint}

In its simplest form, a \hint\ is a list of pointers to \particle s to
be treated by a \stephandler. A handler should then not treat any
other particles in the current \step. The \hint\ may also give a
minimum and maximum scale telling eg.\ a cascade handler to shower
partons starting from one resolution scale, evolving down to the
other.

There are some special \hint s. One is the null hint telling the
handler to figure out for itself what to do given a \step\ object.
Another is the stop \hint, which tells the \collisionhandler\ to stop
the generation in a way such that it can be restarted again by the
controlling program.

It is possible to use inheritance to introduce additional information
in the hints, but a \stephandler\ is not required to take any notice
of this extra information.

\subsection{Initialization}

When an event handler is initialized, the \lumifn\ is initialized and
asked to specify the interval in squared particle-particle rest mass,
$s$ and the type of colliding particles for the trigger collision. If
the interval in $s$ is very large, or if there are other reasons for
several intervals in $s$, the \lumifn\ may specify a list of
intervals. The trigger \collisionhandler\ is then initialized given
the colliding particles and the list of $s$-intervals.

The \collisionhandler\ hands this information to initialize each of
its \subhandler\ objects. Each \subhandler\ then passes this
information on to its \partonextractor\ object which may subdivide the
$s$ intervals further, and for each interval it is asked to specify a
list of intervals in squared parton-parton rest mass, \shat, which it
may produce and a list of possible parton-parton
combinations\footnote{Also different flavour combinations are treated
  separately.} for each \shat\ interval. This information is then
handed the \partonxsec\ objects in the \subhandler, each of which may
subdivide the \shat\ intervals further. Then to each \shat\ interval
and each parton-parton combination the \partonxsec\ objects should
associate a \sigmaxfn\ which should parametrize the maximum
parton-parton cross section the corresponding sub-process may give as
a function of \shat, integrating over all other internal degrees of
freedom. These functions are then passed back to the \partonextractor,
which must convolute them with the parton-parton luminosity function
it represents, to give a maximum integrated cross section for each
sub-process in each $s$-interval.  Hence, for each $s$ interval and
each \shat\ interval and each parton--parton combination and each
\partonxsec\ object, there is now an upper limit of the cross section
stored to be used in the subsequent generation. The \subhandler\ sums
up all these cross sections and hands it back to the
\collisionhandler.

Note that for well-behaved parton-parton cross section and luminosity
functions there will typically be no need for dividing into many
\shat\ intervals. Nor will there in general be any reason to divide up
in several $s$ interval.

The \collisionhandler\ representing the overlayed collisions in an
\eventhandler\ object is then initialized in the same way.

Besides the procedure above, the \eventhandler\ and its
\collisionhandler s also initializes all \stephandler s and
\particledata\ objects to be used to check for internal consistency.

\subsection{Running}

When run, the \eventhandler\ asks its \lumifn\ object to generate an
$s$ which is handed to the trigger \collisionhandler. The
\collisionhandler\ selects a \subhandler\ according to their summed
maximum integrated cross sections for the relevant $s$ interval. This
\subhandler\ then selects an \shat\ interval, a parton-parton
combination and a \partonxsec\ according to the different maximum
cross sections, and asks the \partonextractor\ to generate an \shat\ 
according to this selection. The \partonxsec\ is then asked to
calculate the exact cross section for the selected \shat\ and
parton-parton combination and return the ratio of this cross section
to the previous estimated upper limit. In doing so it may be necessary
for the \partonxsec\ to also generate internal degrees of freedom,
which will be used later in the generation. The corresponding
sub-process is kept with a probability given by the returned ratio
times a corresponding ratio from possible overestimations of the
parton-parton luminosity function in the \partonextractor\ object.

If the selection is discarded the \collisionhandler\ restarts from
selecting a \subhandler.  If the selection is kept, the \partonxsec\
object is asked to generate the exact kinematics of the parton-parton
sub-process in the parton-parton rest frame (a \subcollision\ object).
Given this, the \partonextractor\ is then asked to generate the rest
of the kinematics in the particle-particle rest frame, ie.\ generate
the rapidity of the parton-parton subsystem, boosting it accordingly
to the rest frame of the colliding particles and generate the remnants
of incoming particles, taking care of colour connections in the case
the partons are coloured. Thus we have the initial \step\ object in
the \collision.

The \collisionhandler\ now starts to go through the lists of step
handlers calling each of them in turn, starting with the
post-sub-process handlers, the pre-cascade handlers, the cascade
handler (going through all its hints), the post-cascade handlers etc.
Each of the handlers may throw a \veto\ exception, in which case the
\collisionhandler\ discards the current \collision\ and restarts from
selecting a \subhandler. Each \stephandler\ may also instruct the
\collisionhandler\ to jump back in the list of step handlers to eg.\ 
redo the hadronization steps. This continues until all \stephandler s
have been called, after which the \collisionhandler\ hands back the
generated \collision\ to the \eventhandler.

Note that the \partonextractor\ of the selected subprocess may be
asked by a \stephandler\ to regenerate the particle remnants several
times during the generation of one collision, eg.\ in the case when
the step handler is performing backward parton shower evolution, or
multiple interactions.

The \eventhandler\ boosts the \collision\ to the specified lab frame
and then repeats the procedure with all the handlers corresponding to
overlayed collisions collecting all the obtained \collision s into an
\event, which is handed back to the controlling routine.

In the case of very simple tasks, where the user is not interested in
generating the whole event, it should also be possible to hand the
\eventhandler\ an \event, which is filled \textit{by hand} with eg.\ a
single Z$^0$, and ask it to perform the decay and any subsequent steps
needed.

\subsection{Analysis}

The generated events need to be analyzed in some way. \pyth\ will
contain a basic structure for implementing analysis and will also
provide some standard analysis tasks (see also section
\ref{sec:CLHEPanalysis}). As mentioned above, the analysis may be
implemented as a \stephandler\ and can thus be performed in the middle
of the generation, but may also be applied by the user interface after
the full event has been generated.

\subsection{Finishing}

After a run, a lot of information will have been gathered by the
different handlers. An important example is the Monte-Carlo integrated
cross sections for the different processes that has been generated.
But also other statistics of the run and possible error messages, need
to be communicated to the user. When running in batch mode, the
\eventhandler\ may be told to write out selected information to a log
file, otherwise, the information should be made available through
the user interface.

In addition, the \eventhandler\ may ask each of the handlers which
have been used in a run to write a brief summary of what it has done
and which physical model was used. This should be done in \LaTeX\ 
format and should include references to articles describing the
models. In this way, the log file can be used as a template for
describing the event generation if the result is published,
hopefully ensuring that the simulation is correctly described and that
credit is given to the right people.

\section{Introducing new models in \pyth}

Even if \pyth\ will contain a lot of physics models capable of
simulating a wide range of events for many different experimental
situations, there will always be users who will want to implement
their own physics models for parts of the event generation. Here are
some examples:

\subsection{$n$-fermion generators}

At eg.\ LEP~2, there is a large need for interfacing dedicated
four-fermion generators to standard parton cascade and hadronization
routines. In ref.~\cite{Lep24f} a general form of such a routine was
proposed. Within the \pyth\ framework, this would be accomplished by
writing a class which inherits from \partonxsec, overriding the
virtual methods responsible for communicating the \sigmaxfn\ to the
\partonextractor, and construction of the $n$-fermion kinematics and
colour connections in a \step\ object.

\subsection{Parton density functions}

Today, there exist a standard subroutine package, PDFLIB
\cite{PDFLIB}, which includes most available parton density
parameterizations in the literature, and most event generators are
interfaced to it. For \pyth\ any parameterization can be implemented by
writing a class inheriting from the \partondensity\ class, typically
only overriding the function returning the parton density for a
given parton at a given $x$ and $Q^2$.

\subsection{Other models}

Besides the \partonxsec\ and \partondensity\ classes, it is possible
to introduce new models by writing classes inheriting from the general
\stephandler\ class and the more specialized \cascadehandler,
\mihandler\ and \hadronizationhandler\ classes. Also the
\partonextractor, \remnanthandler, \lumifn, \decayer\ and \decayrater\ 
classes can be inherited from to introduce new models. Finally one can
also imagine inheriting from the \particle\ and \particledata\ classes
to introduce functionality needed by new models.

In all cases \pyth\ will provide well documented example classes which
can be used as templates to simplify the implementation.

\section{User Interface}
\label{sec:UI}

There are many ways in which the user needs to communicate with the
different objects used in the \pyth\ framework. This could of course
be done by writing a controlling program in C++ which directly
interacts with the \pyth\ class library, but the most convenient way
would be to use a high-level (graphical) user interface. \pyth\ will
be distributed together with a rudimentary such interface, but it is
important that the design also allows for the development of more
advanced user interface applications. In particular it is important
that a user implementing a new physical model as eg.\ a \stephandler,
does not need to worry about which user interface is used, but that
there is an easy and standardized way of communicating parameters and
options in the model implementation to any user interface.  For this
reason, \pyth\ will be supply a set of low-level interface classes
which defines the different ways of communication between objects and
a user interface (see also section \ref{sec:geantUI}.)

\subsection{Parameters and Switches}

All classes in \pyth\ which needs to communicate with the user
interface will inherit from the \interface\ class. These includes all
the handler classes, the \lumifn, \partonxsec, \partondensity, etc.

There are four main ways of communicating with \interface\ objects.
There is the setting and retrieving of real-valued \textit{parameters}
and of \textit{switches} with a predefined set of integer-valued
options. These are both central to the way most physics processes are
modeled in an event generator. There is also the \textit{association}
of one or several \interface\ object with another, which is needed
when eg.\ specifying a \partondensity\ object to be used in a
\partonextractor.  Finally there is a more general way of
communication by \textit{commands} in the form of a character string
with an implementation-defined interpretation.

The parameters and switches may be grouped in a hierarchical way in
each class to enable a more logical structure. It is also possible to
define default values and ranges of allowed values. These may be
specified as simple numbers or as functions. The latter can be used
when the limits of one parameter is a function of other parameters,
eg.\ for \cascadehandler s, where the parameter representing the
cutoff scale typically must be larger than the $\Lambda\sub{QCD}$
parameter.

All parameters are local to a given object and are accessed by
specifying the name of the object and the group and name of a
parameter, eg.\\
\hspace*{1cm}\textit{"Object:/Group/Subgroup/Parameter"}.\\
It is also possible to access parameters of one object indirectly if
it is associated to another, eg.\\
\hspace*{1cm}\textit{"Object:Assoc:Parameter"},\\
which accesses the \textit{Parameter} of the object currently pointed
to by the \textit{Assoc} association of \textit{Object}.

To make sure the same parameters for eg.\ hadronization is used
everywhere in the generation one must use the same
\hadronizationhandler\ object everywhere. Otherwise, if one need
different parameter settings for eg.\ overlayed collisions than for
trigger collisions, the \hadronizationhandler\ object may be
duplicated, so that different collision handlers may use different
objects with different parameter settings although they implement the
same physical model.

It is also possible to define truly global parameters as \interface\ 
objects. These could then be assigned to other \interface\ object by
\textit{association}. An example of such an object could be
$\alpha\sub{EM}$, which typically would have a method for retrieving
the value as a function of a given scale.

There may be interdependencies between different objects used in a
simulation. For example, changing the width of the Z$^0$ will change
the cross section used in the \partonxsec\ object describing the
$\qq\rightarrow\mbox{Z}^0$ sub-process. If this is done in mid-run,
the \eventhandler\ object needs to be re-initialized. To avoid
unnecessary re-initialization, there will be a system of
\emph{time-stamps} and \emph{dependency} lists. When the state of an
object is changed in a way that it may influence other objects, its
time-stamp should be updated. This is then noted by the corresponding
\eventhandler\ object, which reinitializes only those objects which
depends on the changed object, in a way similar to that of the Unix
\emph{make} utility.

\subsection{A user session}

As an example of what communication will be necessary, we here outline
what a typical event generator session could look like.
\begin{itemize}
\itemsep -2mm
\item Start the session by reading a standard setup-file (included in
  the \pyth\ distribution) from disk.
\item Copy the the standard \eventhandler\ object for LHC events. This
  would contain a default setup using the default parton densities,
  parton showers, hadronization etc.\ included in \pyth.
\item In the \subhandler\ of the trigger \collisionhandler\ object,
  insert a $\qq\rightarrow\mbox{Z}^0\m{g}$ and a
  $\m{q}\m{g}\rightarrow\m{q}\mbox{Z}^0$ \partonxsec\ object. Then
  replace the \partondensity\ object in the \partonextractor\ with one
  implementing your favorite parton density parameterization.  Finish
  off by modifying the kinematical cuts.
\item In the \particledata\ object representing the Z$^0$, switch off
  all non-leptonic decay modes.
\item Specify the analysis objects to be applied to each \event, the
  number of events to generate, and the files to which statistics and
  analysis results should be written.
\item Save this setup to a file, and start running.
\end{itemize}
All this would typically be done by clicking and dragging within a
graphical user interface. The rudimentary user interface provided by
\pyth, would at least be capable of reading a file with command lines,
which could look something like the example in
fig.~\ref{fig:linemode}.

\begin{figure}
\scriptsize\renewcommand{\baselinestretch}{0.8}
\begin{verbatim}
      read Pythia7Default.conf
      set OutputFile MyLHCRun.out
      copy LHCSample MyLHCRun
      add MyLHCRun:Trigger:SubProcess:PartonXSec q+qbar->Z0+g
      add MyLHCRun:Trigger:SubProcess:PartonXSec q+g->q+Z0
      set MyLHCRun:Trigger:SubProcess:Extractor:PartonDensity MyDensities
      set MyLHCRun:Trigger:Cuts:ptmin 3.0 GeV
      tell Z0 SwitchOff decay all
      tell Z0 SwitchOn decay Z0:?AnyLepton,?AnyAntiLepton;
      add Analysis MyZ0Ptspectrum
      set NumberOfEvents 10000
      save MyLHCRun.conf
      run MyLHCRun
\end{verbatim}

  \caption[dummy]{{\it An example of what an input file to a command-line
      interface could look like for \pyth. Note that the syntax will
      most likely change during the development of \pyth.}}
\label{fig:linemode}
\end{figure}

The user could also write a small main program and access the
different object directly in C++. This could be a bit cumbersome,
however, and typically the user would use a mixture of this and
functions accessing the command-line interface as in the example
program in fig.~\ref{fig:main}.

\begin{figure}
\scriptsize\renewcommand{\baselinestretch}{0.8}
\begin{verbatim}
      #include "Pythia7/main.h"   // The main Pythia include file
                                  // defining all standard things
      #include "MyDensities.h"    // Header file defining the
                                  // MyDensities class inheriting
                                  // fron PartonDensity
      #include "MyZ0Ptspectrum.h" // Header file defining an analysis
                                  // class
      #include <fstream>          // Standard file I/O

      int main() {

        // Create a BaseInterface object to handle the run.
        Pythia::BaseInterface pythia;

        // Read default setup and set output file.
        std::ofstream outfile("MyLHCRun.out");
        pythia.outputFile(outfile);
        pythia.read("Pythia7Default.conf");

        // Copy a standard event handler to be modified.
        pythia.copy("LHCSample", "MyLHCRun");

        // "q+qbar->Z0+g" and "q+g->g+Z0" are standard PartonXSec objects.
        pythia.command("add MyLHCRun:Trigger:SubProcess:PartonXSec q+qbar->Z0+g")
        pythia.command("add MyLHCRun:Trigger:SubProcess:PartonXSec q+g->q+Z0")

        // Create a MyDensity object, name it and add it to the general set
        // of interface objects. Finally tell the parton extractor to use it.
        // GCPtr is a 'smart' reference counting pointer.
        GCPtr<MyDensities> pdf = new MyDensities;
        pdf->name("MyDensities");
        pythia.addInterfaceObject(pdf);
        pythia.command("set MyLHCRun:Trigger:SubProcess:Extractor:PartonDensity"
                       " MyDensities");

        // Finish the setup and save it.
        pythia.command("set MyLHCRun:Trigger:Cuts:ptmin 3.0 GeV");
        pythia.command("tell Z0 SwitchOff decay all");
        pythia.command("tell Z0 SwitchOn decay "
                       "Z0:?AnyLepton,?AnyAntiLepton;");
        pythia.save("MyLHCRun.conf");

        // Create an analysis object.
        MyZ0Ptspectrum analysis;

        // Generate events and analyze them.
        GCPtr<EventHandler> eventHandler = pythia.getEventHandler("MyLHCRun");
        for ( int i = 0; i < 10000; ++i ) {
          GCPtr<Event> event = eventHandler->getEvent();
          analysis.analyze(event);
        }
     
        // Write out statistics and results.
        pythia.writeStatistics();
        analysis.writeHistograms(outfile);

        return 0;
      }
\end{verbatim}

  \caption[dummy]{{\it An example of what a main program using the
      \pyth\ library could look like. Note that the syntax and naming
      of methods will most likely change during the development of
      \pyth.}}
\label{fig:main}
\end{figure}

\section{Error handling}

There are many situations where different kinds of errors may occur
during the event generation. Some of them are due to inconsistencies
which may be detected at initialization time, but others do not occur
until the actual running starts. The error handling in \pyth\ is the
responsibility of the \eventhandler. At initialization time, the
errors are passed directly on to the user interface, for the user to
deal with. When running in batch, all (non-fatal) errors are collected
and summarized to a log file after the run. The errors which can occur
may be more or less severe. \pyth\ will support a number of levels of
severity, ranging from warnings, which do not really affect the
generation, via less serious errors where an event has to be thrown
away, to serious errors where the current run is aborted and very
serious errors where the whole application aborts and dumps core.

The errors are communicated to the \eventhandler\ with simple
member-function calls, except for the cases when an event is discarded
or a run is aborted, in which case the C++ exception mechanism is
used.

\section{Documentation}

A reference manual and a user guide should be written and made
available both on paper and on the web. In addition all header files
should be documented in a way so that they can be automatically
formatted into easily readable pages describing each class and all
public and protected members of each class. As an example, the header
files in CLHEP are written in such a way, and CLHEP includes a very
simple tool called \code{classdoc} to format the header files into
Unix-style \textit{man-pages}. The \code{HepPDT} distribution also
includes a development of \code{classdoc} called \code{h2html}
formatting header files into web pages.

\pyth\ should also include a tool for propagating the documentation of
parameter and switches in physical models implemented within the
framework, both to the reference manual and to the user interface.
This should be done in a way such that a model implementor only needs
to write the documentation of each parameter and option once. This
information would then be extracted and converted both to HTML and
\LaTeX\ code to be inserted in the documentation, and to a format
readable to the user interface.

\section{Relationship to CLHEP/LHC++}

\pyth\ will be built as a class library, and it is clear that many
of its component are general enough to be used also outside of the
program. The general strategy for \pyth\ is that it should depend on
no other software library other than the standard C++ libraries. This
means that a user should be able to obtain the \pyth\ distribution and
install it with the only requirement that he or she has a
standard-conforming C++ compiler with accompanying standard library.
This does not mean, however, that some components of \pyth\ cannot be
shared by other parts of the LHC++ project \cite{LHCXX} especially
CLHEP \cite{CLHEP}. The following is a discussion of such components.

\subsection{The Event record}
\label{sec:CLHEPevent}

Clearly, the event record is something that must be shared with
applications outside of \pyth, since this is the main way of
communicating information about a generated event. The structure
suggested above is however fairy complicated, and although this should
not be a problem for analyzing an event, one could imagine situations,
eg.\ when a program other than \pyth\ is used for the event
generation, where it would be desirable to have a less complicated
event record. The way to solve this potential problem would be to
define a set of standard operations one would need for any kind of
event record, such as \textit{extract a list of all stable charged
  particles}, and either define them in an abstract base class from
which the \pyth\ \event\ and other implementations would inherit, or
in a \emph{traits} class \cite{traits} for event records.

One of the most central among the event record classes is the
\particle\ class. Experience have shown that it is virtually
impossible to agree upon a common particle class for the HEP
community. Not only does the word particle mean different things to
different people (eg.\ generated particle, reconstructed particle,
particle track, etc.), even if there is conceptual agreement of what a
particle is, there are different uses for it, and since it typically
would be a very central class in any application, different people
wants to optimize the implementation in different ways. \pyth\ will,
of course, contain a \particle\ class optimized for event generation,
and it is clear that this cannot be the only particle class the HEP
community. To facilitate the communication between different HEP
applications, it therefore seems reasonable to define a particle
traits class, giving a standard set of operations on a given particle
class and maybe also specifying conversions between different particle
classes.

Another important class is Lorentz and three-vectors. Such classes are
already included in CLHEP, and \pyth\ will, of course, use these. In
addition, \pyth\ may also include a five-vector class inheriting from
the CLHEP Lorentz vector class with an additional mass data member to
allow for optimization of frequently used methods.

\subsection{Particle properties}

Whereas an omni-purpose particle class may be impossible to define, it
should not be that difficult to define a base class describing
standard particle properties. In ref.\ \cite{HepPDT} there is a
suggestion of a set of classes for this purpose. It basically
implements the standard information from the Particle Data Group about
particle properties. The particle property classes in \pyth\ will most
likely be based on these, but will add more information needed to do
event generation as described in section \ref{sec:eventstructure}.

\subsection{Event generator utilities}

Other generator specific classes which should be made general enough
to be used outside of \pyth\ are simple classes such as the templated
\code{Selector} class for efficient selection of objects of a given
class according to given probabilities, and more complicated classes
eg.\ for phase space generation.

\subsection{Analysis}
\label{sec:CLHEPanalysis}

\pyth\ will include a number of tools for performing analysis of
generated events. These will include thrust and sphericity analysis,
jet reconstruction, analysis of energy-energy correlations,
multiplicity distributions etc. These should also be made general
enough to be used outside of \pyth.

\subsection{General utilities}

\pyth\ will contain a number of general utility classes,
which are not necessarily specific to event generation, or indeed to
HEP in general. Because of the requirement that \pyth\ should not
depend on other software libraries, most of these must be included in
the code. Here are some examples:

To reduce the risk of memory leaks, \pyth\ will include a simple
scheme for garbage collection of dynamically allocated objects. This
scheme will consist of a base class for reference counting, together
with a templated \textit{smart pointer} class.

It is conceivable that someone implementing \pyth\ in an
application would like to use it together with an object database,
where eg.\ the produced events would be stored. In this case objects
would be allocated in the database and accessed with \textit{handles}
instead of pointers. It is therefore desirable that the smart pointers
in \pyth\ are parameterized in a way such that they could easily be
modified to encapsulate database handles instead of the simple
reference counted pointers.  This could be achieved by adopting a
strategy like the HepODBMS \cite{HepDB} one, where the use of a
database can be enabled or disabled with a compilation switch.

\pyth\ will use the CLHEP strategy for physical units and constants.

Simple base classes for objects with a name, will be included as well
as classes such as intervals of real numbers etc.

\subsection{Persistency}

Since \pyth\ should not depend on other software libraries, it will
not be tied to an object database. But it will nevertheless be
necessary to store and retrieve objects persistently on
disk. Otherwise it would be much to cumbersome to set up a
\eventhandler\ each run.  \pyth\ will therefore include a simple
scheme for persistency, probably similar to that which previously was
included in CLHEP. Again, it is desirable that this scheme does not
prevent anyone from interfacing \pyth\ to a real object database.

\subsection{User interface}
\label{sec:geantUI}

\pyth\ will not include a fancy user interface, but it is important
that the code is prepared in a way such that implementing a user
interface is easy and streamlined, as explained in section
\ref{sec:UI}. Currently, a low-level set of user-interface classes is
being developed for the Geant 4 project \cite{Geant4}, and it is
likely that \pyth\ will use this strategy or something compatible.

\section{Outlook}

Presently there exist only some fragments of the components of the
basic structure of \pyth. The plan is to now start the actual code
writing, resulting in a proof-of-concept release of \pyth\ in the
summer of 1999, containing the full structure but with only a minimum
of physics models implemented. A complete \pyth\ version, with all the
physics capabilities of the current version of \pythia\ should be
released in the end of 2000.

\section*{Acknowledgments}

This paper is the result of many long and interesting discussions with
Torbjörn Sjöstrand and would not exist without his help.

This work was supported in part by the EU Fourth Framework Programme
`Training and Mobility of Researchers', Network `Quantum
Chromodynamics and the Deep Structure of Elementary Particles',
contract FMRX-CT98-0194 (DG 12 - MIHT).

\appendix
\section*{Appendices}

\section{Coding conventions}

There are no strict coding conventions for \pyth. Especially since
there will initially only be a limited number of authors, there is no
need to enforce any particular coding style.  Below are some
conventions which may be of interest for future users and developers.

\subsection{Name spaces and filenames}

All code in the \pyth\ distribution should be put in the
\emph{namespace} ``\code{Pythia}''.

Header files should only contain declarations and documentation,
preferably with only one class per header file. If the class is called
\code{ClassName}, the header file should be called
\code{ClassName.h}. Inline function definitions should be placed in a
file named \code{ClassName.icc}, which should be \code{\#include}d in
the header file. Non-inline, non-templated functions should be
placed in an implementation file named \code{ClassName.cc},
Non-inline templated function definitions should be placed in a file
name \code{ClassName.tcc}, which could either be \code{\#include}d in
the header file or in the implementation file depending on what the
compiler can handle.  Exception class declarations special to a class
declared in \code{ClassName.h} may be declared in a file named
\code{ClassName.xh} to be \code{\#include}d in the header file which
then can be kept cleaner. To avoid unnecessary header file
inclusions, a forward declaration of a class defined in a header file
may be extracted into a file named \code{ClassName.fh} (cf.\ the
\code{<iosfwd>} header file in the standard C++ library).

\subsection{Garbage collection}

All classes should be reference counted if the corresponding objects
are expected to be dynamically allocated. A base class for reference
counting will be provided together with a templated \textit{smart
  pointer} class to enable rudimentary garbage collection.

\section{Licensing}

The \pyth\ code will be available free of charge to use for any
academic research purpose. The licensing terms will be similar to
those of the non-commercial parts of the LHC++.

\section{Booch notation}
\label{apx:booch}

\begin{figure}[t]
  \begin{center}
        \epsfig{figure=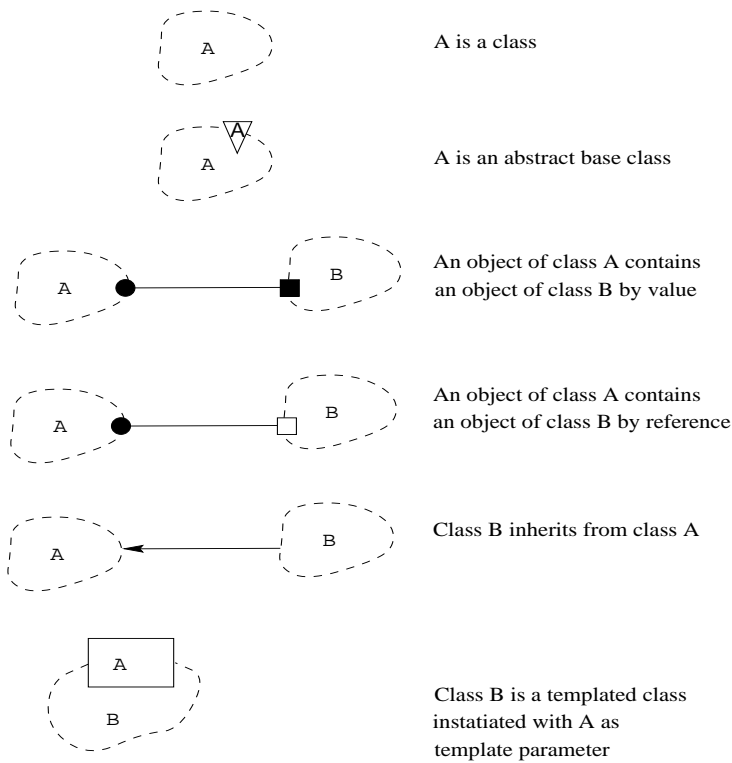,width=10cm,height=10cm}
    \caption[dummy]{{\it Quick guide to the Booch notation used in this paper}}
    \label{fig:booch}
  \end{center}
\end{figure}

In figure \ref{fig:booch} there is an quick guide to the Booch
notation used in the figures in this paper.

\end{document}